\documentclass[12pt]{article}
\usepackage{epsfig,amssymb,amsmath,graphicx,float,latexsym} 

\newcommand{\lsim}{\raisebox{-0.13cm}{~\shortstack{$<$ \\[-0.07cm] $\sim$}}~}
\newcommand{\gsim}{\raisebox{-0.13cm}{~\shortstack{$>$ \\[-0.07cm] $\sim$}}~}

\begin{document}
\renewcommand{\thefootnote}{\fnsymbol{footnote}}

\begin{titlepage}
  
\begin{flushright}
  LAPTH-1325/2009
\end{flushright}

\begin{center}

\vspace{1cm}

{\Large {\bf The Thermal Abundance of Semi--Relativistic Relics}}

\vspace{1cm}

{\bf Manuel Drees}$^{a,\,}$\footnote{drees@th.physik.uni-bonn.de},
{\bf Mitsuru Kakizaki}$^{b,\,}$\footnote{kakizaki@lapp.in2p3.fr}
and 
{\bf Suchita Kulkarni}$^{a,\,}$\footnote{kulkarni@th.physik.uni-bonn.de} \\

\vskip 0.15in
{\it
$^a${Physikalisches Institut and 
Bethe Center for Theoretical Physics, \\ Universit\"at Bonn,
D-53115 Bonn, Germany} \\
$^b${LAPTH, Universit\'e de Savoie, CNRS, \\
B.P. 110, F-74941 Annecy-le-Vieux Cedex, France}
}
\vskip 0.5in

\abstract{Approximate analytical solutions of the Boltzmann equation
  for particles that are either extremely relativistic or
  non--relativistic when they decouple from the thermal bath are well
  established. However, no analytical formula for the relic density of
  particles that are semi--relativistic at decoupling is yet known. We
  propose a new ansatz for the thermal average of the annihilation
  cross sections for such particles, and find a semi--analytical
  treatment for calculating their relic densities.  As examples, we
  consider Majorana-- and Dirac--type neutrinos. We show that such
  semi--relativistic relics cannot be good cold Dark Matter
  candidates.  However, late decays of meta--stable semi--relativistic
  relics might have released a large amount of entropy, thereby
  diluting the density of other, unwanted relics.}

\end{center}
\end{titlepage}
\setcounter{footnote}{0}

\section{Introduction}

The accurate determination of cosmological parameters by up--to--date
observations, most notably by the Wilkinson Microwave Anisotropy Probe
(WMAP) \cite{wmap}, increases the importance of quantitative
predictions. In particular, the estimate of the cosmological relic
abundances of particle species is essential, for the history of the
universe depends on these quantities. One of the most important
examples is the cosmological abundance of dark matter
\cite{kotu,dmrev}, whose mass density is found to be \cite{wmap}
\begin{equation}
  \Omega_{\rm DM} h^2 = 0.1099 \pm 0.0062\, ,
\end{equation}
where $h \simeq 0.7$ is the scaled Hubble parameter in units of
$100~{\rm km~sec}^{-1}~{\rm Mpc}^{-1}$ and $\Omega$ is the mass
density in units of the critical density. 

The abundance of some particle species is determined by solving the
Boltzmann equation, which describes the change of the particle number
caused by particle reactions as well as by the expansion of the
universe \cite{kotu}.  However, there is no analytical general
solution of this nonlinear differential equation, and therefore one
needs to solve the equation numerically in many cases. In early
studies, approximate analytical formulae have been found for the
relativistic \cite{cowsik,kotu} and non-relativistic
\cite{standard,except,improved} regimes.

Stable or long--lived weakly interacting massive particles (WIMPs)
with weak--scale masses are examples of cold relic particles, which
decouple from thermal equilibrium when they are non--relativistic. In
standard cosmology, decoupling of WIMPs occurs in the
radiation--dominated (RD) era after inflation, and analytic
approximate formulae for the WIMP relic abundance have been derived
\cite{standard,except,improved}. In the opposite limit, where
decoupling of particles from the thermal background occurs when they
are relativistic, the relic abundance is approximated by its
equilibrium value at the decoupling temperature, and not sensitive to
details of its freeze--out \cite{kotu}. The resulting relic density
can therefore easily be computed analytically. On the other hand, no
analytical treatment to calculate the relic abundance in the
intermediate regime is known yet.

In this paper, we revisit the relic density of particles $\chi$ that decouple
from the thermal bath when they were semi--relativistic, i.e. at freeze--out
temperature $T_F \sim m_\chi$. Assuming that the Maxwell--Boltzmann
distribution can be used for all participating particles, we introduce an
expression for the thermal average of the $\chi$ annihilation cross section
which smoothly interpolates between the extremely relativistic and
non--relativistic regimes. It is shown that our new ansatz is capable of
reproducing the exact thermally--averaged annihilation cross section with
accuracy of a few percent.  Given this approximated cross section, we can
define the freeze--out temperature by comparing the annihilation rate to the
expansion rate. The assumption that the comoving $\chi$ density remains
constant after freeze--out turns out to be a good approximation for the relic
abundance of semi--relativistic particles.

We also discuss the roles such semi--relativistic particles could play
in realistic cosmological scenarios. It should be emphasized that the
abundance of semi--relativistically decoupled relics tends to be large
because it is only very mildly Boltzmann suppressed. We point out that
scenarios where semi--relativistically decoupling particles form the
Dark Matter have problems with structure formation, Big Bang
Nucleosynthesis (BBN) and/or laboratory measurements. Nevertheless,
such semi--relativistic relics can be useful for diluting the density
of other, unwanted relics by late--time out--of--equilibrium decay
\cite{entropy,late}. As an example, we investigate a scenario of
decaying sterile neutrino that is assumed to depart from thermal
equilibrium when it is semi--relativistic, in sharp contrast to
non--thermal sterile neutrino scenarios \cite{sterile}. Thermal
equilibrium is attained by introducing some higher--dimensional
operator. It is illustrated that an enormous amount of entropy can be
produced without spoiling the successful BBN prediction of the light
element abundances.

This paper is organized as follows: In Sec.~2 we begin by reviewing
briefly the method to calculate relic densities for relativistic and
non--relativistic particles. In Sec.~3 we explain the new formalism
which is applicable for all freeze--out temperatures in case of $S$--
and $P$--wave cross sections. The way to calculate the freeze--out
temperature is also shown. In Sec.~4, the possibility for
semi-relativistic particles to have the observed dark matter relic
density of $\Omega_{\rm DM} h^2 \simeq 0.1$ is considered.  Then, we
discuss the amount of entropy produced by the decay of unstable
semi--relativistic species that decay in less than a second.  Finally,
Sec.~6 is devoted to summary and conclusions. Some properties of
modified Bessel functions are described in Appendix A. In Appendix B
we argue that the use of the Maxwell--Boltzmann distribution in the
definition of the thermally averaged cross section only leads to a
small mistake in the final relic density.

\section{Relic abundances in the non--relativistic and relativistic
  limits}

In this Section we briefly review the standard analytical approximations for
evaluating the relic abundance of hypothetical particles $\chi$
\cite{kotu,standard,improved}. These are applicable to particles that were
either non--relativistic or extremely relativistic at freeze--out.

The number density $n_\chi$ is obtained by solving the corresponding
Boltzmann equation \cite{kotu}. For the moment, we assume that single
$\chi$ production and $\chi$ decay are forbidden by some symmetry or
adequately suppressed. The Boltzmann equation takes a simple form if
one further assumes that the quantum statistics factors describing the
Bose enhancement or Fermi suppression of all final states can be
neglected; in Appendix B it is shown that this is essentially
equivalent to assuming that the distribution functions of all relevant
particles are proportional to the Maxwell--Boltzmann distribution. The
Boltzmann equation for $n_\chi$ can then be written as \cite{kotu}
\begin{equation} \label{eq:boltzmann_n}
  \frac{{\rm d}n_\chi}{{\rm d}t} + 3 H n_\chi =  - \langle \sigma v \rangle
  (n^2_{\chi} - n_{\chi,{\rm eq}}^2)\, ,
\end{equation}
where $n_{\chi,{\rm eq}}$ is the $\chi$ equilibrium number density, $\langle
\sigma v \rangle$ is the thermal average of the annihilation cross section
multiplied by the relative velocity between the two annihilating $\chi$
particles, and $H$ is the Hubble expansion rate of the universe. The second
term on the left--hand side describes the dilution caused by the expansion of
the universe; the first (second) term on the right--hand side decreases
(increases) the number density due to annihilation into (production from)
other particles, which are assumed to be in complete thermal equilibrium.

It is useful to express the above Boltzmann equation in terms of the
dimensionless quantities $Y_\chi = n_\chi/s$ and $Y_{\chi,{\rm eq}} =
n_{\chi,{\rm eq}}/s$. The entropy density is given by $s = (2 \pi^2/45) g_*
T^3$, with $g_*$ being the effective number of relativistic degrees of freedom
and $T$ the temperature of the universe. We also introduce the dimensionless
ratio of $\chi$ mass $m_\chi$ to the temperature, $x = m_\chi/T$. Assuming an
adiabatic expansion of the universe, Eq.(\ref{eq:boltzmann_n}) can then be
rewritten as \cite{kotu}
\begin{eqnarray} \label{eq:boltzmann_y}
  \frac{{\rm d}Y_\chi}{{\rm d}x} = - \frac{\langle \sigma v \rangle s}{H x}
  (Y_\chi^2 - Y_{\chi, {\rm eq}}^2)\, .
\end{eqnarray}

The generic picture of $\chi$ decoupling from the thermal bath is as
follows.  After inflation, the universe becomes radiation--dominated
with expansion rate
\begin{eqnarray} \label{hubble}
  H = \frac{\pi T^2}{M_{\rm Pl}} \sqrt{\frac{g_*}{90}}\, ,
\end{eqnarray}
where $M_{\rm Pl} = 2.4 \times 10^{18}$ GeV is the reduced Planck mass. The
reheat temperate is assumed to be high enough for $\chi$ particles to reach
full thermal (chemical as well as kinetic) equilibrium.\footnote{The case
  where the reheat temperature is too low for $\chi$ to attain chemical
  equilibrium has been discussed in \cite{low,dik1}.}  Thermal equilibrium is
maintained as long as the interaction rate $\Gamma = n_\chi \langle \sigma v
\rangle$ is larger than the Hubble expansion rate $H$. As the temperature
decreases, the interaction rate decreases more rapidly than the expansion rate
does. When the interaction rate falls below the expansion rate, $\chi$ is no
longer kept in thermal equilibrium and the comoving number density $Y_\chi$
becomes essentially constant. This transition temperature is referred to as
the freeze--out temperature $T_F$. 

Analytical expressions for the resulting $\chi$ relic density are known for
the cases where decoupling happens when $\chi$ is non--relativistic $(x_F
\equiv m_\chi / T_F \gg 3)$ or relativistic $(x_F \ll 3)$. We discuss these
two limiting cases in the following Subsections.

\subsection{Relativistic case}

First, consider the case where particles $\chi$ decouple when they are
ultra--relativistic ($x_F \ll 3$).  In this case the equilibrium
number density to entropy ratio $Y_{\chi, {\rm eq}}(x)$ depends on the
temperature only through the number of degrees of freedom $g_*$ of the
thermal bath. Therefore, the final relic abundance is to very good
approximation equal to its equilibrium value at the time of
decoupling:
\begin{eqnarray} 
  Y_{\chi, \infty} \equiv Y_\chi (x \to \infty) 
  = Y_{\chi, {\rm eq}}(x_F) = 0.28~(g_{\rm eff}/g_*(x_F))\, ,
\end{eqnarray}
where
\begin{eqnarray}
  g_{\rm eff} = \left\{
    \begin{array}{l}
      g_\chi \quad \ \ ({\rm for~bosons})\, , \\
      3g_\chi/4 \,({\rm for~fermions})\, ,
    \end{array}
  \right.
\end{eqnarray}
with $g_\chi$ being the number of internal (e.g., spin or color)
degrees of freedom of $\chi$.  Following the conventional notation, we
express the $\chi$ relic density as $\Omega_{\chi} = m_\chi s_0
Y_{\chi,\infty}/\rho_c$, where $\rho_c = 3 H^2_0 M^2_{\rm Pl}$ is the
present critical density of the universe, and $s_0 \simeq 2900~{\rm
  cm}^{-3}$ is the present entropy density.  This yields
\begin{eqnarray}
  \Omega_\chi h^2 = 7.8 \times 10^{-2} \frac{g_{\rm eff}}{g_*(x_F)} 
  \left( \frac{m_\chi}{1~{\rm eV}} \right)\, .
\end{eqnarray}
It should be noted that the relic density is simply proportional to the mass
of the particle in the relativistic case.

\subsection{Non--relativistic case}

For the non--relativistic case where $x_F \gg 3$, the relic abundance strongly
depends on the freeze--out temperature $T_F$ because the equilibrium abundance
$Y_{\chi, {\rm eq}}(x)$ is exponentially suppressed as the temperature
decreases.  The temperature dependence of the thermal average of the
annihilation cross section is obtained using the Taylor expansion in powers of
the velocity squared:
\begin{equation} \label{eq:cross_section}
\langle \sigma v \rangle = a + b \langle v^2 \rangle
  + {\cal O}(\langle v^4 \rangle) =  a + \frac{6 b}{x} 
  + {\cal O}\left( \frac{1}{x^2} \right) \, .
\end{equation}
The numerically--evaluated correct relic abundance is reproduced with an
accuracy of a few percent using the approximate analytic formula
\begin{eqnarray} \label{eq:relic_abundance_y}
  Y_{\chi, \infty} \equiv Y_\chi (x \to \infty)
  = \frac{1}{1.3~ m_\chi M_{\rm Pl}\sqrt{g_*(x_F)}(a/x_F + 3b/x_F^2)}\, .
\end{eqnarray}
For WIMPs with electroweak scale mass, freeze-out occurs at $x_F
\simeq 20$. The corresponding scaled relic density is then given by
\begin{eqnarray} \label{eq:omegah2}
  \Omega_{\chi} h^2 = 2.7 \times 10^8~ Y_{\chi,\infty} 
  \left( \frac{m_\chi}{1~{\rm GeV}} \right)
  = \frac{8.5 \times 10^{-11}~x_F~{\rm GeV}^{-2}}{\sqrt{g_*(x_F)}
    (a + 3 b/x_F)}\, .
\end{eqnarray}
Note that the relic density of a non--relativistic particle is
inversely proportional to its annihilation cross section, but does not
depend explicitly on its mass.

\section{Abundance of semi--relativistically decoupling particles}

In the previous Section, we reviewed the known relativistic and
non--relativistic approximate formulae for the relic abundance. The main aim
of this Section is to find a simple method applicable between the two regimes:
an analytic estimate of the relic abundance of semi--relativistically
decoupling particles ($x_F \sim 3$).

One of the key quantities that determine the freeze--out temperature is the
thermally--averaged cross section. In the non--relativistic case, expressions
for the equilibrium number density as well as for the thermally--averaged
cross section times velocity are rather simple.  For a semi--relativistic
particle, however, the thermally--averaged cross section involves multiple
integrals and cannot be expanded with respect to the velocity nor to the mass.
Here we discuss a method of approximating the thermally-averaged cross
section, by interpolating between its relativistically and
non--relativistically expanded expressions. We employ the Maxwell--Boltzmann
distribution for the equilibrium number density \cite{improved}:
\begin{eqnarray}
  Y_{\chi,{\rm eq}}(x) = \frac{45}{4\pi^4} \frac{g_{\chi} }{g_*(x)}x^2
  K_2(x)\, .
\end{eqnarray}
The thermal average of the cross section is then obtained as \cite{improved}
\begin{eqnarray} \label{sigav}
  \langle \sigma v \rangle = \frac{1}{8 m_\chi^4 T K_2^2 (m_\chi/T)}
  \int_{4 m_\chi^2}^\infty \!\! {\rm d}s \ \sigma (s - 4m_\chi^2) \sqrt{s} \
  K_1 ( \sqrt{s}/T )\, ,
\end{eqnarray}
where $K_1(x)$ and $K_2(x)$ is the first and second modified Bessel function
of the second kind; some properties of these functions are given in Appendix
A.

At first glance the use of the Maxwell--Boltzmann distribution seems improper
for particles that are semi--relativistic at decoupling, let alone for
ultra--relativistic particles. However, we will argue in Appendix B that this
should still yield accurate results for the final relic density. This is
partly due to cancellations between the numerator and denominator of
Eq.(\ref{sigav}), and partly due to the fact that the final result for
$\Omega_\chi$ becomes less sensitive to $x_F$ as $x_F$ decreases.

As examples of the annihilation of particles, we consider the pair
annihilation processes of Dirac and Majorana fermions (e.g. neutrinos)
into a pair of massless fermions, $\chi \bar{\chi} \to f \bar{f}$
\cite{cowsik,lw,other}. It should, however, be noticed that this
assumption includes more general cases of any other species
annihilating from $S-$ or $P-$wave initial states. In a renormalizable
model, the annihilation is mediated by some heavy particle: for
example, a $Z$--boson with tiny coupling with $\chi$, or a new
spin-$1$ boson $U$ \cite{uboson}. We assume that the mass of this
exchange particle is much larger than $m_\chi$, so that the
annihilation amplitude can be described through an effective
four--fermion interaction.

The annihilation of two Dirac fermions proceeds from an $S-$wave, and
the resulting cross section can be parameterized as 
\begin{eqnarray} \label{sigma_S}
  \sigma_S v = \frac{G^2 s}{16\pi}\, ,
\end{eqnarray} 
where $s$ is the center--of--mass energy squared. $G$ denotes the
coupling constant of the four--fermion interaction (e.g. the Fermi
coupling constant, $G_F = 1.17 \times 10^{-5}$ GeV$^{-2}$). Finally, $v$ is
the relative velocity defined as
\begin{eqnarray}
  v = \frac{\sqrt{(p_A \cdot p_B)^2-m_{\chi}^4}}{E_A E_B}\, ,
\end{eqnarray}
where $p_{A,B}$ and $E_{A,B}$ are the four--momenta and energies of the two
incident particles labeled $A$ and $B$. The resulting thermally averaged cross
section is given by
\begin{eqnarray} \label{S_ex}
  \langle\sigma_S v\rangle &=& \frac{G^2}{256 \pi
    m_{\chi}^4TK_2^2(x)}\int_{4m_{\chi}^2}^{\infty}\!\! {\rm d}s\ s^2
  \sqrt{s-4 m_{\chi}^2}K_1 (\sqrt{s}/T)
  \nonumber \\ 
  &=&
  \frac{G^2 m_{\chi}^2}{\pi x^6 K_2^2(x)}\int_0^{\infty}\!\! {\rm d}t\ t^2
  (t^2+x^2)^2 K_1(2\sqrt{t^2+x^2})\, .
\end{eqnarray}
Its relativistic and non--relativistic limits read
\begin{eqnarray}
  \langle \sigma_S v\rangle_{\rm R}
  = \frac{G^2 m_{\chi}^2}{16\pi x^2}(12+5x^2)\, ,
  \quad
  \langle \sigma_S v\rangle_{\rm NR} 
  = \frac{G^2 m_{\chi}^2}{4 \pi}\, ,
\end{eqnarray}
respectively.
A general expression for $\langle \sigma_S v \rangle$ should reproduce these
results for $x \rightarrow 0$ and $x \rightarrow \infty$, respectively. A
simple possibility is
\begin{eqnarray} \label{S_app}
 \langle \sigma_S v\rangle_{\rm app} = \frac{G^2 m_{\chi}^2}{16\pi}\left
 (\frac{12}{x^2}+\frac{5+4x}{1+x}\right )\, .
\end{eqnarray}
It should be noticed that this choice is not unique.

Let us turn to the case of the annihilation from a $P-$wave initial state,
which is e.g. true if $\chi$ is a Majorana fermion. Eq.(\ref{sigma_S}) should
then be replaced by
\begin{eqnarray} \label{eq:majorana}
  \sigma_P v = \frac{G^2 s}{16\pi} \left( 1 - \frac{4m_\chi^2}{s}
  \right)\, .
\end{eqnarray}
Thermal averaging leads to
\begin{eqnarray} \label{P_ex}
  \langle \sigma_P v \rangle &=& \frac{G^2}{256 \pi
    m_{\chi}^4 T K_2^2(x)}\int_{4m_{\chi}^2}^\infty \!\! {\rm d}s\ s (s -
  4m_{\chi}^2)^{3/2}K_1 (\sqrt{s}/T )
  \nonumber \\
  &=&
  \frac{G^2m^2}{\pi x^6 K_2^2(x)}\int_0^\infty \!\! {\rm d}t\ t^4 (t^2+x^2)
  K_1(2\sqrt{t^2+x^2})\, .
\end{eqnarray}
Following the same steps as in the $S-$wave case, we find the following
interpolation:
\begin{eqnarray} \label{P_app}
 \langle \sigma_P v \rangle_{\rm app} = \frac{G^2 m_\chi^2}{16\pi} \left (
 \frac{12}{x^2}+\frac{3+6x}{(1+x)^2}\right )\, .
\end{eqnarray}

\begin{figure}[t!] 
  \begin{center}
    \hspace*{-0.5cm} \scalebox{0.8}{\includegraphics*{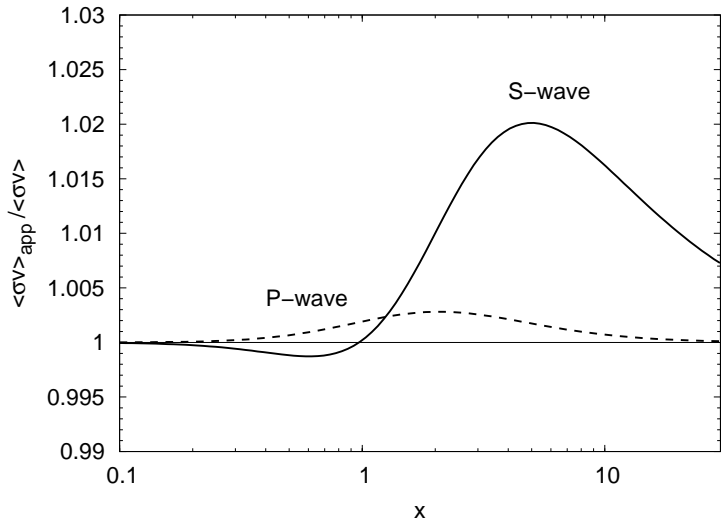}}
    \caption{\footnotesize Ratio of the approximate to the exact
      thermally averaged annihilation cross sections $\langle \sigma v
      \rangle_{\rm app}/\langle \sigma v \rangle$ as a function of
      $x$ for annihilation from an $S-$ (solid line) and $P-$wave
      (dashed) initial state.}
    \label{fig:ratio}
  \end{center}
\end{figure}

Figure \ref{fig:ratio} shows the ratio of the approximate to the exact
cross section $\langle \sigma v \rangle_{\rm app}/\langle \sigma v
\rangle$ for the $S-$ (solid line) and $P-$wave (dashed) cases. Note
that this ratio depends only on $x$.  We see that our approximate
expressions reproduce the exact ones with accuracy of better than 2\%
(0.5\%) for annihilation from an $S-$ ($P-$)wave, even in the
semi--relativistic region ($x \sim 3$).

Using Eqs.(\ref{S_app}) or (\ref{P_app}) instead of the exact
expressions (\ref{S_ex}) or (\ref{P_ex}) greatly reduces the numerical
effort required to solve the Boltzmann equation
(\ref{eq:boltzmann_y}). At the cost of some further loss of accuracy,
an even faster estimate of the relic density can be obtained by using
an approximate solution of the Boltzmann equation instead of the
accurate evaluation of the relic abundance by solving the Boltzmann
equation numerically. The determination of the temperature where some
interaction decouples plays an important role in the analytical
prediction of the relic abundance.  Indeed, for non--relativistically
decoupling particles the relic abundance is sensitive to the
freeze--out temperature $x_F$ because of the Boltzmann suppression. In
this case, a rough estimate of $x_F$ obtained by equalizing the
interaction rate $\Gamma$ and the Hubble expansion rate $H$ is not
sufficient to make an accurate prediction. Here we show that for
semi--relativistically decoupling particles a simple comparison of the
interaction rate and the Hubble expansion rate still gives a
reasonably accurate result for the final $\chi$ relic density.

We define the freeze--out temperature by equalizing the interaction
rate and the Hubble expansion rate,\footnote{This definition of the
  freeze--out temperature should {\em not} be used for the prediction of the
  relic abundance through Eqs.(\ref{eq:relic_abundance_y}) and
  (\ref{eq:omegah2}).}
\begin{eqnarray} \label{freeze}
  \Gamma(x_F) \equiv n_{\chi, {\rm eq}} \langle \sigma v \rangle
  (x_F) = H(x_F)\, .
\end{eqnarray}
We then simply assume that $Y_\chi$ does not change after decoupling from
the thermal bath, so that
\begin{eqnarray} \label{final}
  Y_{\chi,\infty} = Y_{\chi, {\rm eq}} (x_F) \, .
\end{eqnarray}

\begin{figure}[t!] 
  \begin{center}
    \hspace*{-1cm} \scalebox{0.58}{\includegraphics*{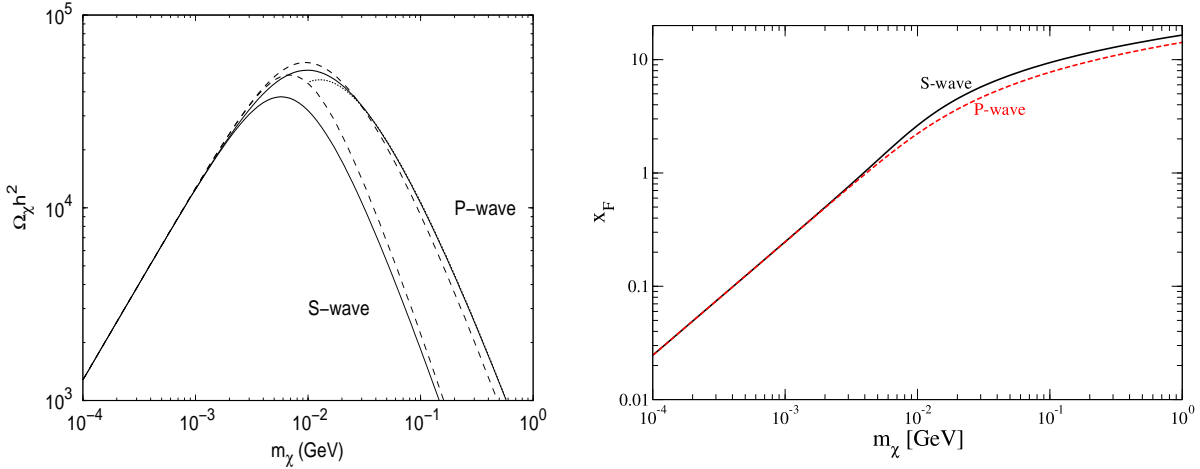}}
    \caption{\footnotesize Scaled freeze--out temperature $x_F$ (right)
      and scaled relic abundance $\Omega_\chi h^2$ (left) as function
      of $m_\chi$.  In the right (left) frame the lower (upper) curves
      are for Majorana fermions, and the upper (lower) curves for
      Dirac fermions. In the left frame the solid curves show exact
      solutions of Eq.(\ref{eq:boltzmann_y}), and the dashed curves our
      analytic approximations. The dotted curve shows the
      non--relativistic approximation, Eq.(\ref{eq:omegah2}), for
      $P-$wave annihilation. Here we take $G = G_F = 1.17 \times
      10^{-5}~\rm{GeV}^2$, $g_* = 10$ and $g_\chi = 2$.}
    \label{fig:neutrino}
  \end{center}
\end{figure}

Let us see to what extent this method can reproduce the correct relic
abundance. As an example, we consider the pair annihilation of
neutrino--like particles via the mediation of the weak SM gauge
bosons\footnote{Our primary concern here is to test our approximation
  for the relic abundance, and thus, as an illustration, we take such
  an unrealistic setup.}. In the left frame of Fig.~\ref{fig:neutrino}
we plot the relic abundance $\Omega_\chi h^2$ as function of
$m_\chi$. The solid curves show predictions for the relic abundance
obtained by solving the Boltzmann equation (\ref{eq:boltzmann_y})
numerically, while the dashed curves have been obtained using the
analytic approximation described above. The upper curves are for
Majorana fermions annihilating from a $P-$wave, and the lower curves
are for Dirac fermions case annihilating from an $S-$wave. We take $G
= G_F = 1.17 \times 10^{-5}~\rm{GeV}^2$, $g_* = 10$ and $g_\chi =
2$\footnote{Since we concentrate on the ratio of the exact to the
  approximate relic densities, we discard the temperature dependence
  of $g_*$, which is basically an overall factor.}. The right frame
shows the corresponding values of $x_F$.

This figure shows that our very simple analytical treatment reproduces
the correct relic density with an error of at most 20\% (5\%) for
semi--relativistically decoupling particles annihilating from an $S-$
($P-$)wave. Not surprisingly, our treatment becomes exact for
particles that are relativistic at decoupling.\footnote{We will see in
  Appendix B that one should use the correct Fermi--Dirac or
  Bose--Einstein expression for $n_{\chi,{\rm eq}}$ in
  Eq.(\ref{final}) in order to accurately predict the relic density
  for $x_F \lsim 1$.}  In the Dirac case, our approximation coincides
with the exact relic abundance for $m_\chi = 1$ GeV, corresponding to
$x_F \simeq 16$; however, the deviation becomes larger again for
larger $m_\chi$. Therefore, our method is not applicable for the
entire region of cold relics. Instead, one could switch to the usual
non--relativistic treatment described in Sec.~2.2 at the cross--over
value, i.e. at $x_F = 16$. In the $P-$wave case the cross--over
already occurs at $m_\chi = 30$ MeV, corresponding to $x_F \simeq
4.5$. The dotted curve shows that the non--relativistic approximation
is already quite reliable at this point. We can thus smoothly match
our approximation to the usual non--relativistic treatment for both
$S-$ and $P-$wave annihilation.

\section{Semi--relativistic dark matter?}
\setcounter{footnote}{0}

As a first application, let us analyze whether semi--relativistically
decoupled particles ($x_F \simeq 3$) can be a dark matter candidate,
whose cosmological abundance should be $\Omega_\chi h^2 \lsim
0.1$. The final number density of such a particle is of order of
$Y_{\chi, {\rm eq}}(x \simeq 3) \sim 10^{-2}$.  Combining this value
with the observed amount of dark matter, the upper bound of the
semi--relativistic particle turns out to be $m_\chi \lsim 100$ eV,
which would thus decouple at temperatures of a few dozen eV. These
particles would therefore still be ultra--relativistic during the
formation of $^4$He. Moreover, the effective coupling would have to be
very large, $G \sim 10^3$ GeV$^{-2}$. Such a scenario is therefore
tightly constrained.

Note that a $\chi$ particle with $m_\chi \sim 100$ eV could
only annihilate into light neutrinos.\footnote{The neutrinos
  themselves aren't in thermal equilibrium with the photons any more
  at $T \lsim 100$ eV. However, as long as $T \gg m_\nu$ they would
  still have a thermal distribution. The Boltzmann equation
  (\ref{eq:boltzmann_n}) should therefore still be applicable, if $T$ is
  taken to be the temperature of the neutrinos, which is somewhat
  lower than the photon temperature.} Moreover, the exchange particle
would also need to be quite light, with mass $\lsim 30$ MeV if all
couplings are $\leq 1$. 

The simplest case is that of a real scalar $\chi$. Since it only adds
a single degree of freedom, its presence would not be in serious
conflict with current BBN constraints \cite{BBN}.  In a renormalizable
theory, $\chi \chi \rightarrow \nu \bar \nu$ could then proceed either
through exchange of a fermion in the $t-$ or $u-$channel, or through
boson exchange in the $s-$channel. 

In the former case, the exchange particle would have to be an $SU(2)$
doublet, if the low--energy theory only contains left--handed, $SU(2)$
doublet neutrinos. The presence of such a light $SU(2)$ doublet
fermion is excluded by LEP data. In principle the light neutrinos
might also be Dirac particles, allowing $\chi \chi \rightarrow \nu_R
\overline{\nu_R}$ annihilation via exchange of a singlet fermion,
possibly $\nu_R$ itself. However, one would then need $\nu_R$ to {\em
  also} be in thermal equilibrium, increasing the number of additional
degrees of freedom present during BBN to an unacceptable level.

For a real scalar $\chi$, $s-$channel exchange could only proceed
through another scalar $\phi$. However, a scalar $\phi$ can only
couple to $\nu_L\overline{\nu_R}$ or its hermitean conjugate. This
scenario would therefore again require $\nu_R$ to have been in
equilibrium. We therefore conclude that such a light $\chi$ particle
cannot be a real scalar.\footnote{Of course, this argument does not
  exclude the possibility that a much heavier real singlet scalar
  $\chi$ could be cold Dark Matter \cite{singlet}.}

If $\chi$ is a complex field, one needs $g_\chi = 2$, which is only
marginally compatible with BBN \cite{BBN}. In principle, $\chi \bar
\chi \rightarrow \nu \bar \nu$ could then proceed through $t-$ or
$u-$channel exchange of an $SU(2)$ singlet. However, then either
$\chi$ or this light exchange particle would have to carry
hypercharge, so that it would have been produced copiously in $Z$
decays. The argument against $s-$channel exchange of a scalar is the
same as for real $\chi$. 

However, a complex $\chi$, either a scalar or a Weyl
fermion\footnote{A massive two--component Weyl fermion can
  equivalently be described by a four--component Majorana fermion.},
could couple to a new gauge boson $U$, which in turn could couple to
$\nu_L \overline{\nu_L}$. This new boson would contribute three
additional bosonic degrees of freedom at $T \gsim m_U$. Since we
already added two degrees of freedom in $\chi$, consistency with BBN
would require $m_U \gsim 1$ MeV. This in turn would require the
coupling of $U$ to left--handed neutrinos to exceed 0.01, assuming its
coupling to $\chi$ is perturbative. By $SU(2)$ invariance, $U$ would
have to couple with equal strength to left--handed charged
lepton. This combination of $U$ boson mass and coupling is excluded,
by a large margin, for both the electron and muon family by
measurements of the respective magnetic moments \cite{uboson}. No
analogous measurement exists for the third generation, so a $U-$boson
with few MeV mass coupling exclusively to third generation leptons and
$\chi$ particles might still be compatible with laboratory
data. However, given that $\mu$ and $\tau$ neutrinos are known to mix
strongly \cite{numix}, a gauge invariant model where $U$ couples to
$\nu_\tau$ but does not couple to muons is difficult, if not
impossible, to construct.

Finally, such a light $\chi$ particle would form hot, or at least
warm, Dark Matter. This possibility is strongly constrained by
observations of early structures in the universe, in particular the
``Lyman$-\alpha$ forest'' \cite{lesgourges}. Such a $\chi$ particle
could thus at best form a sub--dominant component of the total Dark
Matter.

In combination, these arguments strongly indicate that
semi--relativistically decoupling particles should not be absolutely
stable. In the next Section we will show that such particles may
nevertheless have a role to play in the history of the Universe, if
they are metastable.

\section{Entropy production by decaying particles}

In this section we demonstrate that semi--relativistically decoupling
particles can be useful for producing a large amount of entropy, which
could dilute the density of other relics to an acceptable
level. Examples of such relics are decaying gravitinos, which can lead
to problems with Big Bang nucleosynthesis \cite{gravitino}, or
supersymmetric neutralinos, whose relic density often exceeds the
required Dark Matter density by one or two orders of magnitude
\cite{neutralino}. The density of such relics will be diluted only if
the entropy is released after they decouple from the thermal
bath. This will simultaneously dilute any pre--existing baryon
asymmetry. One thus either has to increase the efficiency of early
baryogenesis, or introduce late baryogenesis after the release of the
additional entropy. Both possibilities can be realized in the
framework of Affleck--Dine baryogenesis \cite{AD}.

Generally \cite{entropy,late}, out--of--equilibrium decays of
long--lived particles can only produce a significant amount of entropy
if the decaying particle dominates the energy density of the Universe
prior to its decay. The abundance of non--relativistically decoupling
particles is suppressed by a factor ${\rm e}^{-x_F}$, hence their
contribution to the energy density is small at decoupling. However,
after decoupling their energy density only drops like $R^{-3} \propto
T^3$, while that of the dominant radiation component decreases like
$T^4$ as the Universe cools off. Thermally produced particles can
therefore dominate the energy density of the Universe only at
temperature $T \ll {\rm e}^{-x_F} T_F$. Significant entropy production
by the late decay of nonrelativistically decoupling particles is
therefore only possible if they are simultaneously very massive and
quite long--lived.  For semi--relativistic particles, on the other
hand, the abundance at decoupling is large and thus a significant
amount of entropy can be produced even if their mass is small, since
their density will become dominant quite soon after decoupling.

Let us consider the out--of--equilibrium decay of long--lived particles
which semi--relativistically decoupled from the thermal
background. For simplicity we work in the instantaneous decay
approximation, i.e. we assume that all $\chi$ particles decay at time
$t_d = \tau_\chi$, where $\tau_\chi$ is the lifetime of $\chi$. While
this approximation does not describe the time dependence of the
entropy (or temperature) for $t \sim \tau_\chi$ very well, it does
reproduce the entropy enhancement factor, i.e. the entropy at $t \gg
\tau_\chi$, quite accurately. We assume that $\chi$ particles were in
full thermal equilibrium for sufficiently high temperatures in the RD
epoch.  When the temperature decreased to $T = T_F \simeq m_\chi$, the
$\chi$ number density $n_\chi$ froze out. At decoupling, $\chi$
particles contributed a few percent to the total energy density of the
universe; however, as noted earlier, the ratio of the radiation and
$\chi$ energy densities decreased by a factor $T_d / T_F =
\sqrt{t_F /\tau_\chi}$ between decoupling and decay of $\chi$; here
$T_d$ refers to the temperature at time $t = \tau_\chi$, just prior to
$\chi$ decay. If $\tau_\chi \gg t_F$, the $\chi$ energy density at the
time of the $\chi$ decay is well approximated by $\rho_{\chi,d} =
m_\chi n_{\chi,d}$, and dominated over the radiation. In this case,
the ratio of the final entropy density $s_f$ after the decay to the initial
entropy density $s_i$ before the decay is given by \cite{entropy}
\begin{eqnarray} \label{eq:entropy_release}
   \frac{s_f}{s_i} = 0.82~g_*^{1/4} 
     \frac{m_\chi Y_{\chi,d} \tau_\chi^{1/2}}{M_{\rm Pl}^{1/2}} \, ,  
\end{eqnarray}
for $s_i \ll s_f$.  Here $Y_{\chi, d} = n_{\chi,d} / s_i$ is
proportional to the $\chi$ abundance just prior to its decay.

In the light of the BBN prediction of the primordial abundances of the
light elements, the $\chi$ lifetime is constrained as $\tau_\chi \lsim
1~{\rm sec}$ \cite{trmin}. Equations (\ref{eq:omegah2}) and
(\ref{eq:entropy_release}) show that the entropy ratio is proportional
to the relic density $\Omega_\chi h^2$ that $\chi$ would have if it
were stable. We saw in Fig.~2 that for fixed coupling $G$ this
quantity is maximal if $T_F \sim m_\chi$; more accurately, the maximum
of $\Omega_\chi h^2$ is achieved for $x_F = 1.8 \ (2.1)$ if $\chi$
particles annihilate from an $S-$ ($P-$)wave initial state. Entropy
production by late $\chi$ decays is thus most efficient when the
$\chi$ particles decoupled semi--relativistically, with their lifetime
fixed to the maximal value of $\sim 1$ sec.

We can construct a feasible scenario that fulfills these conditions by
introducing a sterile neutrino which mixes with an ordinary neutrino.
Here we treat both $m_\chi$ and the mixing angle $\theta$ as free
parameters. In sharp contrast to conventional cosmological scenarios
with sterile neutrinos \cite{sterile}, the sterile neutrino is assumed
to be in thermal equilibrium in the early universe.  In ordinary
sterile neutrino models, thermal equilibrium is not reached because
the Yukawa coupling of the sterile neutrino with SM particles is
tiny. One possible method for the $\chi$ pair production and
annihilation to reach thermal equilibrium is to extend sterile
neutrino models by adding another hypothetical boson $Z'$.  Let it
have coupling $g_{Z'f}$ with the SM fermion pair $f\bar{f}$, and $g_{Z'\chi}$
with the sterile neutrino pair. If the $Z'$-boson mass $m_{Z'}$ is
larger than the $\chi$ energy, the $\chi$ annihilation cross section
has the form of Eq.(\ref{eq:majorana}) with $G = g_{Z'\chi} g_{Z'f}/m_{Z'}^2$.
Although $g_{Z'f}$ and $m_{Z'}$ are constrained by high energy
experiments, $g_{Z'\chi}$ can be as large as unity. Therefore, $\chi$
annihilation can be in thermal equilibrium before its
semi--relativistic decoupling. Decoupling occurred at $T \sim m_\chi$
if $m_\chi \simeq 1 \ {\rm GeV} \cdot \left( 3 \cdot 10^{-9} \ {\rm
    GeV}^{-2} / G \right)^{2/3} g_*^{1/6}$.

In order to estimate the amount of entropy released by the decay of
sterile neutrinos in this setup, we have to calculate their
lifetime. For simplicity we ignore propagator effects. When the
sterile neutrino mass is smaller than the $W-$boson mass $m_W = 80$
GeV, it decays into three SM fermions, with decay width
\begin{eqnarray}
 \Gamma_\chi =   
\left( 27 - 16 \sin^2 \theta_W + \frac{80}{3} \sin^4 \theta_W
 \right)
 \frac{G_F^2 m_\chi^5}{192 \pi^3} \sin^2 \theta\, ,
\end{eqnarray}
where $\sin^2 \theta_W = 0.23$ is the weak mixing angle.  When the
sterile neutrino mass is larger than the $Z-$boson mass $m_Z = 91$
GeV, the sterile neutrino predominantly decays into a SM gauge boson
and a lepton. Its decay width is then proportional to $m_\chi^3$, and
given by
\begin{eqnarray}
 \Gamma_\chi =
 \frac{G_F m_\chi^3}{8 \sqrt{2} \pi} \left[
 2 \left( 1 - \frac{m_W^2}{m_\chi^2} \right)^2
 \left( 1 + \frac{2 m_W^2}{m_\chi^2} \right) 
 + \left( 1 - \frac{m_Z^2}{m_\chi^2} \right)^2
 \left( 1 + \frac{2 m_Z^2}{m_\chi^2} \right) 
 \right] \sin^2 \theta\, .
\end{eqnarray}
In the in--between case where $m_W < m_\chi < m_Z$, we obtain
\begin{eqnarray}
 \Gamma_\chi & = &
 2 \frac{G_F m_\chi^3}{8 \sqrt{2} \pi} 
 \left( 1 - \frac{m_W^2}{m_\chi^2} \right)^2
 \left( 1 + \frac{2 m_W^2}{m_\chi^2} \right) \sin^2 \theta
 \nonumber \\
 &+&  \left( 11 - 20 \sin^2 \theta_W + \frac{80}{3} \sin^4 \theta_W \right)
 \frac{G_F^2 m_\chi^5}{192 \pi^3} \sin^2 \theta\, .
\end{eqnarray}

\begin{figure}[t!]
  \begin{center}
    \hspace*{-0.5cm} \scalebox{1}{\includegraphics*{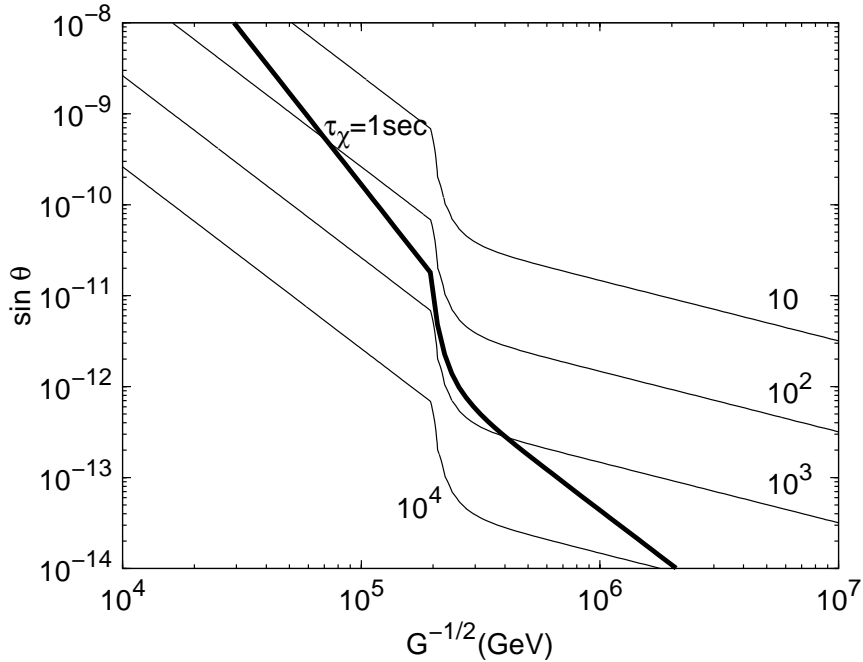}}
    \caption{\footnotesize Contour of the entropy increase $s_f/s_i$ 
      caused by semi--relativistic sterile neutrino decay in
      the $(1/\sqrt{G}, \sin \theta)$ plane. We choose $m_\chi$ such
      that $x_F=2.1$. The solid line indicates the BBN limit on the
      sterile neutrino lifetime $\tau_\chi = 1$ sec.}
    \label{fig:sterile}
  \end{center}
\end{figure}

Figure \ref{fig:sterile} shows contours of the entropy increase
$s_f/s_i$ due to sterile neutrino decay in the $(1/\sqrt{G}, \sin
\theta)$ plane.  We set the freeze--out temperature to $x_F = 2.1$,
which maximizes $m_\chi Y_{\chi, i}$; this can be achieved by chosing
the mass $m_\chi$ appropriately. The thick line indicates the BBN
limit on the sterile neutrino lifetime, $\tau_\chi = 1$
sec. Eq.(\ref{eq:entropy_release}) shows that for given neutrino mass,
the released entropy will be maximal if $\theta$ is chosen such that
$\tau$ reaches this upper limit.

The behavior of the contours in Fig.~\ref{fig:sterile} is easy to
understand from Eq.(\ref{eq:entropy_release}). In the relevant limit
$\theta \ll 1$ and keeping $g_*$ constant, we have $\tau_\chi \propto
\theta^{-2} m_\chi^{-5} \ (\theta^{-2}m_\chi^{-3})$ for $m_\chi < \ (>)
\ m_W$. The entropy ratio thus scales as $\theta^{-1} m_\chi^{-3/2}
\propto \theta^{-1} G \ (\theta^{-1} m_\chi^{-1/2} \propto \theta^{-1}
G^{1/3})$ for $m_\chi < \ (>) \ m_W$. Along the $\tau_\chi = 1$ sec
contour, the entropy release increases proportional to $m_\chi \propto
G^{-2/3}$ both for $m_\chi < m_W$ and for $m_\chi > m_W$.
Fig.~\ref{fig:sterile} can be extended to even smaller $G$,
i.e. larger $Z'$ masses, so long as $m_\chi$ is smaller than the
re--heat temperature after inflation, so that $\chi$ was in thermal
equilibrium in the RD epoch. If at the same time $\theta$ is decreased
so that $\tau_\chi = 1$~sec remains constant, very large entropy
dilution factors could be realized,
\begin{equation} \label{rat_max}
\frac {s_f} {s_i} \leq 10^3 \cdot \left( \frac {G^{-1/2}} {10^6 \ {\rm
      GeV}} \right)^{4/3}\,.
\end{equation}

This result is only valid if the mixing--induced interactions of
$\chi$ are not in thermal equilibrium for $T \lsim m_\chi$. Since
these interactions are also responsible for $\chi$ decay, this
assumption is satisfied whenever $\tau_\chi \gg t_F$; we saw in the
discussion of Eq.(\ref{eq:entropy_release}) that this strong
inequality is in any case a condition for significant entropy release
from $\chi$ decay.

We finally note that for given $m_\chi$ the entropy released in $\chi$
decays is maximal if $G$ is so small that $\chi$ was
ultra--relativistic at decoupling, since this maximizes $Y_{\chi,{\rm
    eq}}(x_F)$. Again setting $\tau_\chi = 1$ sec by appropriate
choice of $\theta$, this yields
\begin{equation} \label{rat_max1}
\frac {s_f} {s_i} \leq 10^4 \cdot \frac {m_\chi} {10^3 \ {\rm
      GeV}} \,.
\end{equation}

\section{Conclusion}

In this paper we have developed an approximate analytic method for
calculating the thermally--averaged annihilation cross section of
semi--relativistically decoupling particles and for estimating their
relic density. We have shown that this approximate solution can be
smoothly matched to the well--known non--relativistic approximation at
the point of intersection. We have argued that such relics cannot form
the observed cosmological dark matter.  However, we pointed out that
the late decay of metastable semi--relativistically decoupling relics
can be an efficient source of entropy production. As an example of
this entropy production mechanism we discussed a scenario with a
sterile neutrino, and illustrated to what extent entropy can be
increased.

\section*{Acknowledgments}

This work was partially supported by the Marie Curie Training Research
Network ``UniverseNet'' under contract no.  MRTN-CT-2006-035863, and
by the European Network of Theoretical Astroparticle Physics ENTApP
ILIAS/N6 under contract no.  RII3-CT-2004-506222.  The work of
M.K. was also supported by the Marie Curie Training Research Network
``HEPTools'' under contract no.  MRTN-CT-2006-035505.

\appendix

\section*{Appendix A: Modified Bessel Functions}

In this appendix, we summarize some properties of the modified Bessel
function.  Using an integral representation, the modified Bessel
function of the second kind is defined by
\begin{eqnarray}
  K_\nu(z) = \frac{\sqrt{\pi} (z/2)^\nu}{\Gamma(\nu + 1/2)} \int^\infty_1
  \!\! {\rm d}t \ {\rm e}^{-zt}(t^2 - 1)^{\nu - 1/2}, \quad
  {\rm Re}(\nu) > - \frac{1}{2}, \ {\rm Re}(z) >0\, .
\end{eqnarray}
In particular, the calculation of the relic abundance involves
$K_1(z)$ and $K_2(z)$,
\begin{eqnarray}
  K_1(z) & = & z \int^\infty_1
  \!\! {\rm d}t \ {\rm e}^{-zt}(t^2 - 1)^{1/2}, \quad
  {\rm Re}(z) >0\, ,
  \nonumber \\
  K_2(z) & = & \frac{z^2}{3} \int^\infty_1
  \!\! {\rm d}t \ {\rm e}^{-zt}(t^2 - 1)^{3/2}, \quad
  {\rm Re}(z) >0\, .
\end{eqnarray}
The lower order terms of the series expansion of $K_1(z)$ and $K_2(z)$
are given by
\begin{eqnarray}
  K_1 (z) & = & \frac{1}{z} + \cdots \, , 
  \nonumber \\
  K_2 (z) & = & \frac{2}{z^2} - \frac{1}{2} + \cdots \, .
\end{eqnarray}
The asymptotic expansion of $K_\nu(z)$ is given by
\begin{eqnarray}
  K_\nu(z) \sim \sqrt{\frac{\pi}{2z}} {\rm e}^{-z} 
  \left( 1 + \frac{4 \nu^2 - 1}{8z} + \cdots \right)\, .
\end{eqnarray}

\section*{Appendix B: Validity of the Maxwell--Boltzmann Distribution}

In the calculations of this paper we used the Maxwell--Boltzmann (MB)
distribution also for particles that were semi--relativistic at
decoupling; this assumption is e.g. implicit in Eq.(\ref{sigav}). At
first sight this seems quite dangerous. For example, at $T = m_\chi$,
i.e. $x=1$, the MB result for $n_{\chi,{\rm eq}}$ overestimates the
Fermi--Dirac distribution by about 7\%, and underestimates the
Bose--Einstein distribution by about 10\%. Since $\chi$ annihilation
always involves two $\chi$ particles, one might assume that the total
error associated with the use of the MB distribution is about twice as
large. In this Appendix we show that the MB distribution can indeed be
used to compute the thermally averaged cross section and the
decoupling temperature as long as $x_F \gsim 1$. For smaller $x_F$,
one has to use the proper Fermi--Dirac or Bose--Einstein distribution
only in the very last step, when calculating $Y_{\chi,\infty}$. 

We begin by expanding the true distribution function,
\begin{equation} \label{expand_f}
f_{\chi,{\rm eq}}(E_\chi) = \frac {1} {{\rm e}^{E_\chi/T} \pm 1 }
\simeq {\rm e}^{- E_\chi/T} \left( 1 \mp {\rm e}^{-E_\chi/T} \right)
\, ,
\end{equation}
where the upper (lower) sign is for fermionic (bosonic) $\chi$
particles. Note that the correction term in parentheses has exactly
the same form as the ``statistics factors'' appearing in the collision
term of the full Boltzmann equation \cite{kotu}. For consistency these
statistics factors therefore also have to be included. Up to first
order in these correction factors, the temperature dependent terms in
the integrand defining the collision term for $\chi \chi
\leftrightarrow f \bar f$ processes then read for fermionic $f$:
\begin{eqnarray} \label{collterm}
{\cal I} &=& {\rm e}^{-(E_{\chi_1} + E_{\chi_2}) / T} \cdot \left[
c_\chi^2 \left( 1 \mp {\rm e}^{-E_{\chi_1}/T} \mp {\rm
    e}^{-E_{\chi_2}/T} - {\rm e}^{-E_f/T} - {\rm e}^{-E_{\bar f}/T}
\right)
\right. \nonumber \\  && \left. \hspace*{2.5cm}
- \left( 1 - {\rm e}^{-E_f/T} - {\rm e}^{-E_{\bar f}/T} \mp c_\chi
  {\rm e}^{-E_{\chi_1}/T} \mp c_\chi {\rm e}^{-E_{\chi_2}/T} \right)
\right]
\nonumber \\
&=& {\rm e}^{-(E_{\chi_1} + E_{\chi_2}) / T} \cdot \left[
\left( c_\chi^2 - 1 \right) 
\left( 1 \mp {\rm e}^{-E_{\chi_1}/T} \mp {\rm
    e}^{-E_{\chi_2}/T} - {\rm e}^{-E_f/T} - {\rm e}^{-E_{\bar f}/T}
\right)
\right. \nonumber \\  && \left. \hspace*{2.5cm}
\pm \left( c_\chi - 1 \right) \left( {\rm e}^{-E_{\chi_1}/T} + {\rm
    e}^{-E_{\chi_2}/T} \right) \right] \,;
\end{eqnarray}
here $c_\chi = f_\chi / f_{\chi,{\rm eq}}$ is independent of energy as
long as $\chi$ is in kinetic equilibrium (through elastic scattering
on SM particles); in that case we can equivalently write $c_\chi =
n_\chi / n_{\chi,{\rm eq}}$. In order to derive the full collision
term, ${\cal I}$ has to be multiplied with the squared matrix element
and integrated over phase space \cite{kotu}.

In the usual treatment of WIMP decoupling, all the exponential terms
in the square parentheses are neglected, so that the collision term
becomes proportional to $n_\chi^2 - n_{\chi,{\rm eq}}^2$ times the
thermally averaged cross section defined in
Eq.(\ref{sigav}). Unfortunately the full correction term introduces
additional dependence on the final state energies $E_f$ and $E_{\bar
  f}$. In order to keep the numerics manageable, we assume that they
can be replaced by $E_{\chi_1}$ and $E_{\chi_2}$, respectively. This
is certainly true (by energy conservation) for the sum $E_f + E_{\bar
  f}$; this has already been used in deriving
Eq.(\ref{collterm}). Note furthermore that we'll need the collision
term for temperatures $\gsim T_F$, where $|c_\chi - 1| \ll 1$, so that
we can approximate $c_\chi - 1 \simeq (c_\chi^2-1)/2$. These
approximations yield
\begin{equation} \label{collterm_app}
{\cal I} \simeq \left( c_\chi^2 - 1 \right) {\rm e}^{-(E_{\chi_1} +
  E_{\chi_2}) / T} \left[ 1 - \kappa \left({\rm e}^{-E_{\chi_1}/T} +
    {\rm e}^{-E_{\chi_2}/T} \right) \right]\,,
\end{equation}
where $\kappa = 1/2$ (3/2) for bosonic (fermionic) $\chi$
particles. In the following we will assume $\chi$ to be fermionic,
which according to Eq.(\ref{collterm_app}) should lead to larger
deviations from the MB result. 

Inserting this corrected collision term into the Boltzmann equation,
and following the formalism of \cite{improved}, finally yields a
modified thermally averaged cross section times initial state
velocity:
\begin{eqnarray} \label{sigav_mod}
\langle \sigma v \rangle &=& \frac {1} {n_{\chi,{\rm eq}}^2} \frac
{g_\chi^2} {8 (2\pi)^4} \int d E_+ d E_- ds (\sigma \cdot F)(s) {\rm
  e}^{-E_+/T} \ \ \ \ \ \ \ \ \ \ \ \ \ \ \ \ \ \ \ \ \ \ \ \ \ \ \ \
\ \ \
\nonumber \\ 
&& \hspace*{2.5cm} \cdot \left[ 1 - \kappa \left( {\rm e}^{-(E_+ +
      E_-)/(2T)} + {\rm e}^{-(E_+ - E_-)/(2T)} \right) \right]\,,
\end{eqnarray}
with $F = 2 s \sqrt{1 - 4 m_\chi^2/s}$, $E_+ = E_{\chi_1} +
E_{\chi_2}$ and $E_- = E_{\chi_1} - E_{\chi_2}$. This reduces to
Eq.(\ref{sigav}) if the expression in square parentheses is simply
replaced by 1. In the following we assume that $\chi$ particles
annihilate from an $S-$wave. $P-$wave annihilation would favor larger
energies, where the correction terms in Eq.(\ref{sigav_mod}) are
smaller. Note that we also have to use the expanded form
(\ref{expand_f}) of the distribution function when calculating
$n_{\chi,{\rm eq}}$ in Eq.(\ref{sigav_mod}); otherwise the solution of
the Boltzmann equation will not yield $n_\chi \simeq n_{\chi,{\rm
    eq}}$, including the correction terms, at $T \gg T_F$.

\begin{figure}[h!]
\begin{center}
\rotatebox{270}{\scalebox{0.5}{\includegraphics*{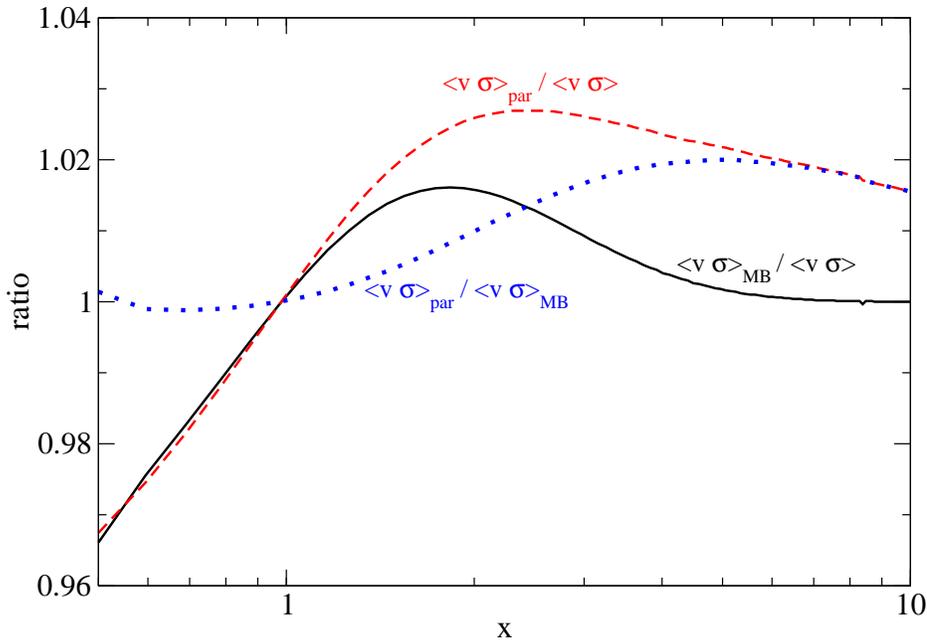}}}
\caption{\footnotesize Various approximations for the thermally
  averaged cross section as function of the scaled inverse temperature
  $x = m_\chi / T$ for fermionic particles annihilating from an
  $S-$wave. The solid (black) curve shows the ratio of the corrected
  cross section (\ref{sigav_mod}) to the Maxwell--Boltzmann (MB) result
  (\ref{sigav}), while the dashed (red) curve shows this ratio if
  Eq.(\ref{sigav}) is replaced by our approximation (\ref{S_app}). The
  dotted (blue) curve is the same as the solid curve in Fig.~1.}
\label{app_1}
\end{center}
\end{figure}

The size of the correction terms in Eq.(\ref{sigav_mod}) is
illustrated by the solid (black) curve in Fig.~\ref{app_1}. We see
that the correction amounts to less than 2\% for all $x \gsim 1$. This
is due to a strong cancellation between the corrections in the
integrand of Eq.(\ref{sigav_mod}) and those in the overall factor
$1/n_{\chi,{\rm eq}}^2$. The dashed (red) curve shows that for $x \sim
2$ the errors due to the use of the MB distribution and due to our
simple parameterization (\ref{S_app}) add up, leading to a total error
of about 2.7\% at most. The Fermi--Dirac corrections to the thermally
averaged cross section begin to be significant for $x \lsim
0.5$. However, here one enters the ultrarelativistic regime, where the
final relic density is no longer sensitive to the decoupling
temperature. We therefore expect the effect of using the MB
distribution in Eq.(\ref{sigav}) on the final prediction of the relic
density to be quite small throughout.

\begin{figure}[h!]
\begin{center}
\rotatebox{270}{\scalebox{0.5}{\includegraphics*{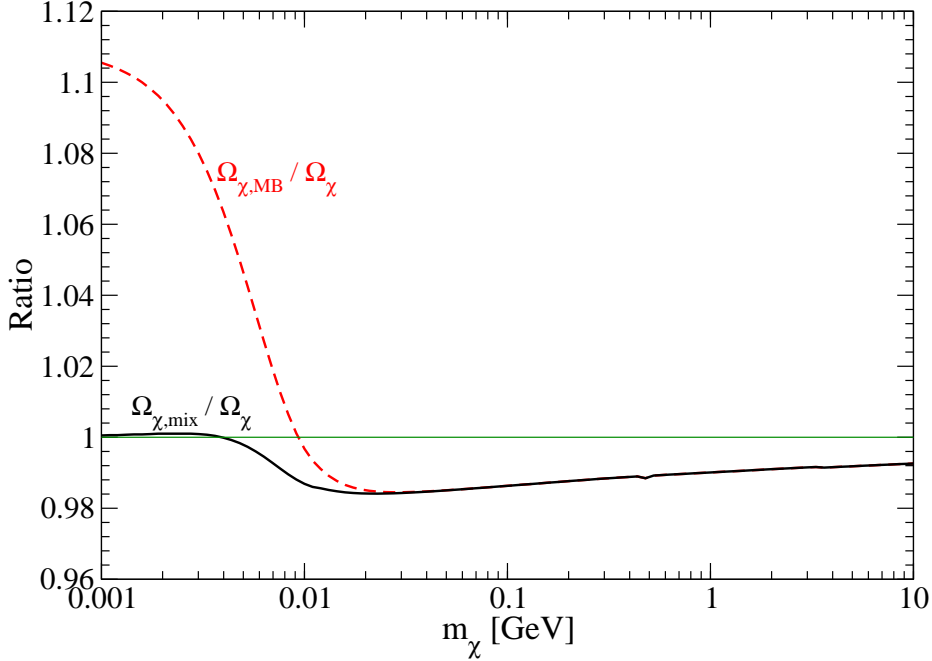}}}
\caption{\footnotesize Effect of using the Maxwell--Boltzmann
  distribution on the predicted relic density, calculated using the
  approximation $Y_{\chi,\infty} = Y_{\chi,{\rm eq}}(x_F)$, for
  fermionic $\chi$ particles annihilating from an $S-$wave initial
  state. The dashed (red) curve shows the ratio of the prediction
  using the Maxwell--Boltzmann distribution everywhere to the
  corrected prediction based on Eq.(\ref{expand_f}) and
  (\ref{sigav_mod}). The solid (black) curve shows the analogous
  ratio, where correct Fermi--Dirac distribution has been used to
  evaluate $Y_{\chi,{\rm eq}}(x_F)$, but $\langle \sigma v \rangle$ and
  $x_F$ have still been obtained using the MB distribution. Parameters
  are as in Fig.~2.}
\label{app_2}
\end{center}
\end{figure}

This is illustrated in Fig.~\ref{app_2}, where the relic density has
been calculated using the simple assumption $Y_{\chi,\infty} =
Y_{\chi,{\rm eq}}(x_F)$; we have used the same parameters as in
Fig.~2. The dashed (red) curve shows that using the MB distribution
everywhere will overestimate the relic density for $m_\chi \lsim 5$
MeV, i.e. for $x_F \lsim 1$. However, the black curve shows that this
can easily be corrected by using the Fermi--Dirac distribution {\em
  only} in the final step, i.e. when calculating $Y_{\chi,{\rm
    eq}}(x_F)$; $\langle \sigma v \rangle$ and $x_F$ can still been
calculated using the MB distribution. This validates our treatment in
the main text.

\end{document}